\documentclass{ws-ijmpb}
\usepackage{amsmath}
\usepackage{onlinecite}

\makeatletter

\newcommand\@pnumwidth{1.55em}
\newcommand\@tocrmarg{2.55em}
\newcommand\@dotsep{4.5}
\def\contentsname{Contents}
\newcommand\tableofcontents{%
    \section*{\contentsname
        \@mkboth{%
           \MakeUppercase\contentsname}{\MakeUppercase\contentsname}}%
    \@starttoc{toc}%
    }
\newcommand*\l@part[2]{%
  \ifnum \c@tocdepth >-2\relax
    \addpenalty\@secpenalty
    \addvspace{2.25em \@plus\p@}%
    \setlength\@tempdima{3em}%
    \begingroup
      \parindent \z@ \rightskip \@pnumwidth
      \parfillskip -\@pnumwidth
      {\leavevmode
       \large \bfseries #1\hfil \hb@xt@\@pnumwidth{\hss #2}}\par
       \nobreak
       \if@compatibility
         \global\@nobreaktrue
         \everypar{\global\@nobreakfalse\everypar{}}%
      \fi
    \endgroup
  \fi}
\newcommand*\l@section[2]{%
  \ifnum \c@tocdepth >\z@
    \addpenalty\@secpenalty
    \addvspace{1.0em \@plus\p@}%
    \setlength\@tempdima{1.5em}%
    \begingroup
      \parindent \z@ \rightskip \@pnumwidth
      \parfillskip -\@pnumwidth
      \leavevmode \bfseries
      \advance\leftskip\@tempdima
      \hskip -\leftskip
      #1\nobreak\hfil \nobreak\hb@xt@\@pnumwidth{\hss #2}\par
    \endgroup
  \fi}
\newcommand*\l@subsection{\@dottedtocline{2}{1.5em}{2.3em}}
\newcommand*\l@subsubsection{\@dottedtocline{3}{3.8em}{3.2em}}
\newcommand*\l@paragraph{\@dottedtocline{4}{7.0em}{4.1em}}
\newcommand*\l@subparagraph{\@dottedtocline{5}{10em}{5em}}

\let\citeonline\citelow

\makeatother

\numberwithin{equation}{section}

\newcommand{\br}{{\bf r}}

\newcommand{\kB}{k_{\mathrm{B}}}

\def\pacs#1{\par
    \vspace*{8pt}
    {\authorfont{\leftskip18pt\rightskip\leftskip
    \noindent{PACS numbers}\/:\ #1\par}}\par}
    
\begin{document}

\markboth{N. H. March, G. G. N. Angilella, and R. Pucci}{Phase changes via solvable models}

%
\catchline{}{}{}{}{}
%

\title{\uppercase{REVIEW\\ \medskip 
Insights gained from solvable models into a variety of phase transitions, including emergent assemblies plus isoelectronic series of atomic ions}}

\author{N. H. MARCH}

\address{Department of Physics, University of Antwerp,\\
Groenenborgerlaan, 171, B-2020 Antwerp, Belgium\\[.5\baselineskip]
Oxford University, Oxford, UK}

\author{G. G. N. ANGILELLA\footnote{Corresponding author. E-mail:
giuseppe.angilella@ct.infn.it}}

\address{Dipartimento di Fisica e Astronomia, Universit\`a di Catania,\\
Via S. Sofia, 64, I-95123 Catania, Italy\\[.5\baselineskip]
Scuola Superiore di Catania, Universit\`a di Catania,\\ Via
Valdisavoia, 9, I-95123 Catania, Italy\\[.5\baselineskip]
CNISM, UdR Catania, Via S. Sofia, 64, I-95123 Catania, Italy\\[.5\baselineskip]
INFN, Sez. Catania, Via S. Sofia, 64, I-95123 Catania, Italy}

\author{R. PUCCI}

\address{Dipartimento di Fisica e Astronomia, Universit\`a di Catania,\\
Via S. Sofia, 64, I-95123 Catania, Italy\\[.5\baselineskip]
CNISM, UdR Catania, Via S. Sofia, 64, I-95123 Catania, Italy}

\maketitle

\begin{history}
\received{1 October 2014}
\accepted{6 October 2014}
\end{history}

\begin{abstract}
Three solvable models are set out in some detail in reviewing different types of phase transitions. Two of these relate directly to emergent critical phenomena, \emph{viz.} melting and  magnetic transitions in heavy rare-earth metals, and secondly, via the $3d$ Ising model, to critical behaviour in an insulating ferromagnet such as CrBr$_3$. The final `transition', however, concerns ionization of an electron in an isoelectronic series with $N$ electrons as the atomic number $Z$ is reduced below that of the neutral atom. These solvable models are, throughout, brought into contact either with experiment, or with very precise numerical modelling on real materials.
\end{abstract}




\tableofcontents

\markboth{N. H. March, G. G. N. Angilella, and R. Pucci}{Phase changes via solvable models}

\clearpage

\section{Background and outline}

Though there has been major theoretical progress with respect to critical behaviour in spin magnets and in the vicinity of the liquid-vapour critical point (see \emph{e.g.} Refs.~\citeonline{Kadanoff:00,Kardar:07}), other areas of phase transitions remain relatively crudely treated by genuine many-body methods.

Therefore, in this review, we have appealed, almost at once, to analytically solvable models in order to gain further insight into a variety of phase transitions. The first of these models resorts to the simplification of pair potentials, which are widely useful for treating both rare gas solids and a variety of simple metals. However, we use it by way of an introduction to the material presented in section~\ref{sec:re} below on melting of some heavy rare-earth metals. Then, we exploit recent work using the embedded atom model based on density functional theory (DFT) \cite{SteinleNeumann:99} to treat various regularities associated with such heavy rare-earth metals.

Section~\ref{sec:CrBr3} then provides a brief introduction to critical phenomena, which leads into a review of recent detailed theoretical treatment of the critical exponents in the three-dimensional Ising model (see \emph{e.g.} Ref.~\citeonline{Kardar:07} for further introductory material). Section~\ref{sec:iso} then treats a very different third type of phase transition, \emph{viz.} that occurring in isoelectronic series of atomic ions having $N$ electrons with nuclei of charge $Ze$. As $Z$ is lowered, for given $N$, below the neutral atomic number, to a critical (usually non-integral) value $Z_c (N)$, one electron ionizes and this is frequently termed in the literature as a `phase transition'. But, of course, it is of quite different character from the two cooperative phase transitions treated in section~\ref{sec:re} immediately below, and also in the vicinity of a critical point in section~\ref{sec:CrBr3}.

\section{Relation between melting and magnetic phase transition temperatures in six heavy rare-earth metals: insights from a solvable model}
\label{sec:re}

We start by considering the correlation between both melting temperature $T_m$ and N\'eel temperature $T_{\mathrm N}$ with the de~Gennes factor $\xi$ in six rare-earth materials, as studied by Ayuela and March \cite{Ayuela:14}. The de~Gennes factor characterizes $f$ localized electrons in rare-earth metals, and is defined as $\xi = (g-1)^2 J(J+1)$, with $g$ the Land\'e factor, and $J$ the eigenvalue of the total angular momentum \cite{Jensen:91}.

\begin{table}[t]
\tbl{Values of the de~Gennes factor $\xi$ and melting temperatures $T_m$
for six heavy rare earth metals \protect\cite{Chikazumi:97}. Also shown is the theoretically predicted
vacancy formation energy \protect\cite{SteinleNeumann:99} $E_v$, and the ratio $E_v /\kB T_m$. 
The last two columns finally report the bulk modulus, $B$, and the ratio $\frac{1}{2} B\Omega / \kB T_m$,
discussed in the main text, $\Omega$ being the atomic volume.}{%
\begin{tabular}{cr@{.}lcr@{.}lr@{.}lr@{.}lr@{.}l}
\hline
Material & \multicolumn{2}{c}{$\xi$} & $T_m$ (K) &
\multicolumn{2}{c}{$E_v$ (eV)} & \multicolumn{2}{c}{$E_v / \kB T_m$} 
& \multicolumn{2}{c}{$B$} & \multicolumn{2}{c}{$\frac{1}{2} B\Omega / \kB T_m$} \\
Tm & 1 & 17 & 1820 & \multicolumn{2}{c}{---} & \multicolumn{2}{c}{---} & \multicolumn{2}{c}{---} & \multicolumn{2}{c}{---} \\
Er & 2 & 55 & 1795 & 1 & 31 & 10 & 2 & 44 & 6 & 24 & 0 \\
Ho & 4 & 50 & 1745 & 1 & 27 & 10 & 2 & 40 & 9 & 23 & 3 \\
Dy & 7 & 08 & 1682 & 1 & 2 & 10 & 1 & 40 & 5 & 23 & 9\\
Tb & 10 & 50 & 1632 & 1 & 18 & 10 & 1 & 38 & 9 & 23 & 9\\
Gd & 15 & 75 & 1587 & 1 & 14 & 10 & 0 & 37 & 8 & 24 & 3\\
\hline
\end{tabular}%
}
\label{tab:xiTm}
\end{table}

Table~\ref{tab:xiTm} first lists the six heavy rare-earth metals Tm to Gd, to be discussed in some detail below. The second column reports the corresponding numerical values of the above de~Gennes factor $\xi$ for these materials \cite{Chikazumi:97}. Already, in the book of Chikazumi and Graham \cite{Chikazumi:97} the N\'eel temperature $T_{\mathrm{N}}$ is shown to correlate well with $\xi$ for the first five crystals in Table~\ref{tab:xiTm}, Gd having solely a ferromagnetic phase. This matter of both N\'eel and Curie ($T_{\mathrm C}$) temperatures will be discussed in some detail in section~\ref{ssec:deGexp} below.

However, we will begin with the melting temperatures $T_m$, the experimental values of which are recorded in the third row of Table~\ref{tab:xiTm}. Before discussing the correlation of these temperatures with the de~Gennes factor, however, attention will be focussed immediately below on a solvable model which relates $T_m$ to the vacancy formation energy $E_v$ in an admittedly simplistic pair potential model, which is most appropriate in fact to the \emph{crystalline} rare gas solid Ar.

\subsection{Solvable model of Bhatia and March relating vacancy formation energy $E_v$ at melting point to melting temperature $T_m$ plus features of liquid structure}

Early work of Faber \cite{Faber:72} gave a ${\bf k}$-space formula for the vacancy formation energy $E_v$ at the melting point in terms of the liquid structure factor $S(q)$. This works best for close-packing in `simple' metals like Pb, with fcc structure. Indeed, the vacancy formation energy in this solvable model neglects relaxation around the vacant site: an inappropriate assumption for, say, the bcc alkali metals \cite{Flores:81,March:13}. Within Faber's \cite{Faber:72} ${\bf k}$-space theory, the vacancy formation energy $E_v$ is related to the pair potential $\phi(r)$ in real space as \cite{Minchin:74}
\begin{equation}
E_v = -\frac{\rho}{2} \int g(r) \phi(r) d\br - \frac{\rho}{6} \int r \frac{\partial \phi}{\partial r} g(r) d\br ,
\label{eq:Ev}
\end{equation}
where $g(r)$ is the radial distribution function, and $\rho$ denotes the atomic number density. The radial corrrelation function $g(r)$ can be related in turn to the measured structure factor $S(q)$, which is defined as the Fourier transform of $g(r)$. Bhatia and March \cite{Bhatia:84} then took results available on liquid argon near $T_m$ to estimate the first contribution to the right hand side of Eq.~(\ref{eq:Ev}) as
\begin{equation}
-\frac{\rho}{2} \int g(r) \phi(r) d\br \simeq 0.08~\mbox{eV} ,
\end{equation}
which is already in semiquantitative accord with the measured vacancy formation energy $E_v$ for Ar, an insulating material, near freezing.

This prompted Bhatia and March \cite{Bhatia:84} to relate the Ornstein-Zernike (OZ) correlation function \cite{March:02} $c(r)$ to $E_v$, via the $\br$-space form Eq.~(\ref{eq:Ev}) quoted above. Its Fourier transform $\tilde{c}(q)$ is related to $S(q)$ by
\begin{equation}
\tilde{c}(q) = \frac{S(q)-1}{S(q)} ,
\label{eq:cS}
\end{equation}
which readily follows from the $\br$-space OZ theory \cite{March:02}. The merit of $c(r)$ is that, in pair potential theory, it is closely connected to $\phi(r)$, within classical statistical mechanics (appropriate to the nuclei even in liquid metals) by
\begin{equation}
c(r) \mapsto - \frac{\phi(r)}{\kB T} .
\label{eq:cconnect}
\end{equation}
Using an established statistical mechanics approximation for potentials (as is the case for liquid metals) having hard cores of diameter $\sigma$, say, one chooses $c(r)$ as in Eq.~(\ref{eq:cconnect}) for $r>\sigma$, and as $c_{\mathrm{hard\mbox{-}core}}(r)$ for $r<\sigma$. Bhatia and March \cite{Bhatia:84} invoked Percus-Yevick \cite{Percus:58} analytic theory for $c_{\mathrm{hard-core}}$. Then, a straightforward calculation yields \cite{Bhatia:84}
\begin{equation}
\frac{E_v}{\kB T_m} \simeq \frac{1}{2} \left[ \tilde{c}(q=0) - c(r=0) -3 \right]_{T=T_m} .
\label{eq:BM}
\end{equation}
As shown from X-ray plus neutron scattering experiments by Bernasconi and March \cite{Bernasconi:86}, one has $c(r=0)\approx -40$ for most liquid metals near freezing. Also, in view of Eq.~(\ref{eq:cS}), we have
\begin{equation}
\tilde{c}(0) = 1 - \frac{1}{S(0)} .
\end{equation}
But from statistical mechanics fluctuation theory, it is well-known \cite{March:02} that $S(0)$ is precisely given by
\begin{equation}
S(0) = \rho \kB T \kappa_T = \frac{\kB T}{B\Omega} ,
\label{eq:S0}
\end{equation}
where $\rho$ is again the atomic number density, $\kappa_T$ is the isothermal compressibility, $B=\kappa_T^{-1}$ is the bulk modulus, and $\Omega$ denotes the atomic volume. This yields from Eq.~(\ref{eq:BM}) for many simple liquids near freezing that \cite{Bhatia:84}
\begin{equation}
\frac{E_v}{\kB T_m} + \frac{1}{2} \frac{B\Omega}{\kB T_m} \approx 20 .
\label{eq:approx20}
\end{equation}
Ayuela and March \cite{Ayuela:14} recognized that heavy rare-earth metals would involve many-body correlations, so they used experimental values for $\kB T_m$ and $B\Omega$ at $T=T_m$, while for $E_v$ they employed existing results from the embedded atom density functional theory (DFT) model \cite{SteinleNeumann:99}. Their interest, as remarked at the outset, was to see to what regularities one was led by correlating experiments at first on $E_v$ and $\kB T_m$ with the de~Gennes factor $\xi$. This is reviewed in Table~\ref{tab:xiTm}. One immediately recognizes that the ratio $E_v /\kB T_m$ turns out to be almost constant, $E_v /\kB T_m \approx 10$, as is indeed often assumed by material scientists. Also the ratio $B\Omega/\kB T_m$, entering Bhatia and March's formula involving the pair potential above, remains pretty constant also within the rare-earth metals considered by Ayuela and March \cite{Ayuela:14}, where many-body correlations are expected to play a significant role. However, the linear combination of the two ratios on the left-hand side of Eq.~(\ref{eq:approx20}), as well as the numerical value on the right-hand side, are importantly altered by transcending pair potentials, as is done by use of the embedded atom model for the rare earths \cite{SteinleNeumann:99}.

\subsection{Experimental measurements on heavy rare-earth metals correlated by de~Gennes factor}
\label{ssec:deGexp}

\begin{figure}[t]
\centering
\includegraphics[height=0.8\columnwidth,angle=-90]{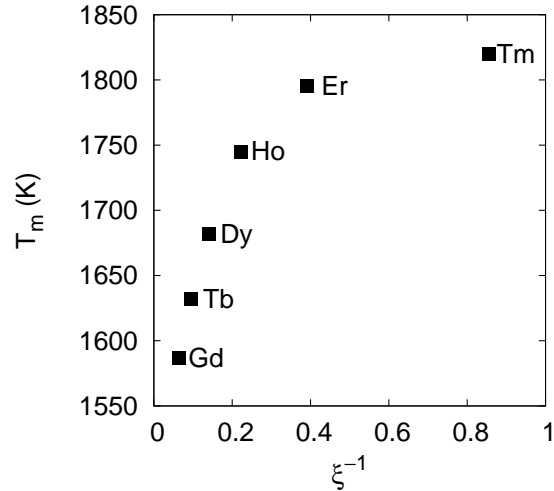}
\caption{Melting temperatures $T_m$ as a function of the inverse de~Gennes factor $\xi^{-1}$, for the six heavy rare-earth metals under consideration in section~\ref{sec:re}. Redrawn after Ref.~\protect\cite{Ayuela:14}.}
\label{fig:a1}
\end{figure}

\begin{figure}[t]
\centering
\includegraphics[height=0.8\columnwidth,angle=-90]{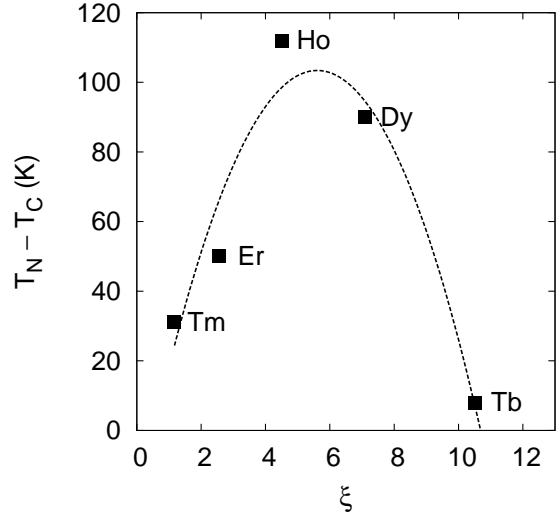}
\caption{Shows the non-monotonic dependence of the difference between the N\'eel and the Curie temperatures, $T_{\mathrm{N}}-T_{\mathrm{C}}$, for the six heavy rare-earth metals under consideration in section~\ref{sec:re}. Dashed line is a guide to the eye. Redrawn after Ref.~\protect\cite{Ayuela:14}.}
\label{fig:a2}
\end{figure}

To complete this section, we want to make close contact between experimental measurements on heavy rare-earth metals pertaining to melting and magnetic phase transitions and the de~Gennes factor $\xi$. Fig.~\ref{fig:a1} first shows measured melting temperatures $T_m$ for the six heavy rare-earth metals plotted vs $\xi^{-1}$. There is a remarkably smooth correlation in evidence. Fig.~\ref{fig:a2} next turns to magnetic phase transitions. In particular, this figure shows how the difference between N\'eel and Curie temperatures behaves as a function of the de~Gennes factor. Again, though now with some scatter, as Ayuela and March \cite{Ayuela:14} note, there is again a clear correlation. 

\begin{figure}[t]
\centering
\includegraphics[height=0.8\columnwidth,angle=-90]{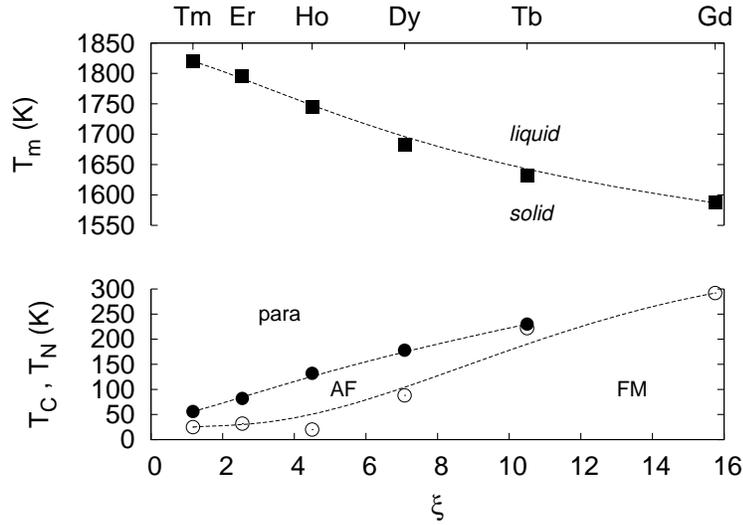}
\caption{`Unconventional' phase diagram for the six heavy rare-earth metals under consideration in section~\ref{sec:re}. Along with the liquid and solid phases, separated by the melting line $T_m = T_m (\xi)$, with $\xi$ the de~Gennes factor, magnetic phases are also marked as paramagnetic (para), antiferromagnetic (AF) below the N\'eel temperature $T_{\mathrm{N}}$, and ferromagnetic (FM) below the Curie temperature $T_{\mathrm{C}}$. Dashed lines are guides to the eye. Redrawn after Ref.~\protect\cite{Ayuela:14}.}
\label{fig:a3}
\end{figure}

But the most important plot these authors made was in the form of what they term an `unconventional' phase diagram, appropriate at atmospheric pressure. This is shown in Fig.~\ref{fig:a3}. The change of scale has to be stressed, as one moves from the melting temperature curve to the considerably lower magnetic transition temperatures. We stress again here the important fact noted already in the book by Chikazumi and Graham \cite{Chikazumi:97} that $T_{\mathrm{N}} \varpropto \xi^{2/3}$. Ayuela and March \cite{Ayuela:14} best fitted the data with
\begin{equation}
T_{\mathrm{N}} = A \xi^{2/3} ,
\label{eq:Axi}
\end{equation}
with $A\simeq 48$~K. Of course, it remains of considerable interest for the future to give a first principles theory of the established semiempirical Eq.~(\ref{eq:Axi}).

\subsection{Future directions}
\label{ssec:CrBr3:fd}

Matthai, Ayuela, and March are currently trying to generalize the previous considerations to embrace $3d$ transition metals, such as Fe, Ni, and Co. Unfortunately, no clear cut characterization of these $3d$ elements exists paralleling the de~Gennes factor $\xi$ for the rare earths. A promising approach seems to be that of including the Gr\"uneisen parameter $\gamma_{\mathrm{G}}$. This would not be entirely surprising, since $\gamma_{\mathrm{G}}$ is a measure of phonon anharmonicity, and vacancy equilibrium concentration at a given temperature $T$ is expected to depend on this parameter.

\section{CrBr$_3$ near the ferromagnetic transition: Generalization of Zhang's prediction for the critical exponents of the 3$d$ Ising model if $\alpha\neq0$}
\label{sec:CrBr3}

Very early work on the liquid--vapour ($\ell-v$) critical point was that of Ornstein and Zernike \cite{Ornstein:14,Ornstein:18}. Indeed, it was in this context that they defined the now-called direct correlation function $c(r)$, which we already introduced in section~\ref{sec:re} above. If we write $g(r)-1 \equiv h(r)$, the asymptotic form they proposed for the quantity $h(r)$, often termed the total correlation function in liquid state theory, was
\begin{equation}
h(r) \varpropto r^{-1} \exp(-r/\xi_{\mathrm{OZ}} ) 
\end{equation}
where $\xi_{\mathrm{OZ}} \varpropto [1-\tilde{c}(0)]^{-1/2} = S(0)^{1/2}$ has the meaning of a correlation length between density fluctuations. This length diverges as the critical point is approached, as $S(0)$ tends to infinity, since the isothermal compressibility $\kappa_T$ diverges there [see Eq.~(\ref{eq:S0}) above].

While the above results are useful first approximations, the modern theory of criticality (see \emph{e.g.} Ref.~\citeonline{Fisher:67}) requires two independent exponents to describe the long-range correlations between density fluctuations. Thus, near the critical point the total correlation function has the asymptotic behaviour
\begin{equation}
h(r) \varpropto r^{2-d-\eta} \exp(-r /\xi_c ) ,
\label{eq:h}
\end{equation}
$d$ being the dimensionality (here and below, $d=3$; for the exact Onsager solution of the $d=2$ Ising model, see \emph{e.g.} Ref.~\citeonline{Reichl:09}). The correlation length $\xi_c$ in Eq.~(\ref{eq:h}) takes the form $\xi_c \varpropto t^{-\nu}$, where $t=|T-T_c|/T_c$ is the reduced temperature, measuring the distance in temperature from the critical point at $T=T_c$. At the same time, $\kappa_T$ at the critical density $\rho_c$ is $\kappa_T \varpropto t^{-\gamma}$, and the difference between liquid and gas densities $\rho_\ell -\rho_v \varpropto t^\beta$.

In addition to the four critical exponents thereby introduced above, we note next that the definitions of the exponents which characterize the singular behaviour of physical properties near the critical point are most frequently introduced using a language appropriate to spin magnetism. In addition to the four exponents already defined above ($\kappa_T$ becoming the susceptibility $\chi$, and $\rho_\ell -\rho_v$ the magnetization $M$), two further exponents $\alpha$ and $\delta$ are introduced. These quantify the behaviour of the heat capacity in zero field, $C \varpropto t^{-\alpha}$, and of the order parameter as a function of magnetic field $H$ at $t=0$, $M\varpropto H^{1/\delta}$.

This is the point at which we choose, as an example of an experimental system, the insulating ferromagnet CrBr$_3$ (Refs.~\citeonline{Ho:69,Schofield:69}). We do not want to get closely involved in whether this system CrBr$_3$ is really appropriately described by the Ising model (the main theme here), which remains possibly controversial. But in the above references experimental values are recorded for the critical exponents $\gamma$, $\beta$, and $\delta$, which relate to information contained in the $3d$ Ising model table given below (Tab.~\ref{tab:Zhang}). With this as background, we turn to the explicit way Tab.~\ref{tab:Zhang} was constructed \cite{March:14,March:14a} for the $3d$ Ising model.

The crucial combinations of critical exponents set out in Ref.~\citeonline{March:14} have been used by one of us \cite{March:14a} to derive relations reducing to Zhang's results \cite{Zhang:07a} for the $3d$ Ising model when the critical exponent $\alpha$ is set equal to zero. The basic assumption underlying these relations is that Zhang's prediction $\gamma=5/4$ for the $3d$ Ising case is exact. This is made since Wilson \cite{Wilson:72}, in his pioneering work on the $\epsilon$-expansion, quoted $\gamma=1.244$, while using higher-order terms plus Borel summability, Kardar \cite{Kardar:07} in his book cites $\gamma=1.238$. No experiments in the foreseeable future will be able to distinguish between these three values, and therefore Zhang's explicit prediction $\gamma=5/4$ will be retained as the starting point in the review that follows. In a previous account \cite{March:14}, it was shown that with dimensionality $d$:
\begin{equation}
d=\frac{(2-\alpha)(2-\eta)}{\gamma}
\label{eq:Zhang}
\end{equation}
which, as this combination of (three) critical exponents yielded precisely the dimension $d$, was referred to as a `crucial combination'.

Using also the Josephson \cite{Josephson:67a} relation in the form
\begin{equation}
d=\frac{2-\alpha}{\nu} ,
\label{eq:Josephson}
\end{equation}
it follows from Eq.~(\ref{eq:Zhang}) that
\begin{equation}
\gamma\nu^{-1} = 2-\eta .
\label{eq:Zhang3}
\end{equation}
This relation, in fact, holds for the $d$-dimensional Ising model independently of the value of the critical exponent $\alpha$.

\subsection{Refinement of Zhang's critical exponents for $d=3$ if $\alpha\neq0$}

Using the Rushbrooke \cite{Rushbrooke:63} relation, one has
\begin{equation}
\alpha + 2\beta + \gamma =2
\label{eq:Rushbrooke}
\end{equation}
and therefore for $d=3$ with $\gamma=5/4$, Eq.~(\ref{eq:Rushbrooke}) yields the result
\begin{equation}
\beta = \frac{3}{8} - \frac{\alpha}{2} .
\label{eq:Zhang2}
\end{equation}
For reference throughout this section, Zhang's predicted values of the six critical exponents under discussion are recorded in Table~\ref{tab:Zhang}. When $\alpha$ is put equal to zero, Eq.~(\ref{eq:Zhang2}) reduces to $\beta=3/8$, also listed in Table~\ref{tab:Zhang}.

\begin{table}[t]
\tbl{Zhang's \protect\cite{Zhang:07a} predicted values for critical exponents of the $3d$ Ising model.}{%
\begin{tabular}{cccccc}
\hline
$\alpha$ & $\beta$ & $\gamma$ & $\delta$ & $\eta$ & $\nu$ \\
0 & $3/8$ & $5/4$ & $13/3$ & $1/8$ & $2/3$ \\
\hline
\end{tabular}%
}
\label{tab:Zhang}
\end{table}

Josephson's relation above, Eq.~(\ref{eq:Josephson}), yields, for $d=3$, that
\begin{equation}
\nu = \frac{2-\alpha}{3} .
\label{eq:Josephson2}
\end{equation}
Turning to the exponent $\delta$, the Hubbard-Schofield result \cite{Hubbard:72,Zhang:13,March:13a} will be utilized. This reads
\begin{equation}
\delta = \frac{d + (2-\eta)}{d - (2-\eta)} .
\label{eq:HubbardSchofield}
\end{equation}
Invoking Eq.~(\ref{eq:Zhang3}), this becomes
\begin{equation}
\delta = \frac{d + \gamma\nu^{-1}}{d - \gamma\nu^{-1}} .
\end{equation}
To find $\eta$, Eq.~(\ref{eq:Zhang}) with $d=3$ and $\gamma=5/4$ gives
\begin{equation}
(2-\eta) = \frac{15}{4(2-\alpha)} ,
\end{equation}
or
\begin{equation}
\eta = \frac{1-8\alpha}{8-4\alpha} .
\label{eq:Zhang4}
\end{equation}
Using again $\gamma=5/4$ for $d=3$, Eq.~(\ref{eq:HubbardSchofield}) takes the form
\begin{equation}
\delta = \frac{3\nu + \frac{5}{4}}{3\nu - \frac{5}{4}} .
\end{equation}
This result leads to the formula below, when $\nu$ is related to $\alpha$ by Eq.~(\ref{eq:Josephson2}), \emph{viz.}
\begin{equation}
\delta = \frac{13-4\alpha}{3-4\alpha} .
\label{eq:Zhang5}
\end{equation}
Eqs.~(\ref{eq:Zhang2}), (\ref{eq:Josephson2}), (\ref{eq:Zhang4}), and (\ref{eq:Zhang5}) immediately give back Zhang's predictions recorded in Table~\ref{tab:Zhang}, when $\alpha=0$. Since these are the central equations of this section of the present review, they have been collected in Table~\ref{tab:Zhang2}.

\begin{table}[t]
\tbl{Formulae generalizing Zhang's \protect\cite{Zhang:07a} predictions for critical exponents in the $3d$ Ising model, recorded
in Table~\protect\ref{tab:Zhang}, given $\gamma=5/4$ and $\alpha\neq0$. One may also note that the combination $\frac{3}{2} \nu-\beta = \frac{5}{8}$, independently of $\alpha$, for $d=3$. Also $\eta\nu = \frac{4}{3}\beta-\frac{5}{12}$.}{%
\begin{tabular}{cccl}
\hline
 & & & \\
\hbox to 3truecm{\hfill} & $\displaystyle\beta = \frac{3}{8} - \frac{\alpha}{2}$ & \hbox to 3truecm{\hfill} & Eq.~(\ref{eq:Zhang2}) \\
 & & & \\
\hbox to 3truecm{\hfill} & $\displaystyle\delta = \frac{13-4\alpha}{3-4\alpha}$ & \hbox to 3truecm{\hfill} & Eq.~(\ref{eq:Zhang5}) \\
 & & & \\
\hbox to 3truecm{\hfill} & $\displaystyle\eta = \frac{1-8\alpha}{8-4\alpha}$ & \hbox to 3truecm{\hfill} & Eq.~(\ref{eq:Zhang4}) \\
 & & & \\
\hbox to 3truecm{\hfill} & $\displaystyle\nu = \frac{2-\alpha}{3}$ & \hbox to 3truecm{\hfill} & Eq.~(\ref{eq:Josephson2}) \\
 & & & \\
\hline
\end{tabular}%
}
\label{tab:Zhang2}
\end{table}

\subsection{The $\epsilon$-expansion revisited}

To proceed further, however, we return to the $\epsilon$-expansion. Wilson \cite{Wilson:72}, as well as giving $\gamma=1.244$ mentioned above, also estimated for $d=3$ that the critical exponent $\eta$ was 0.037, which was in fact quoted in the Abstract of his Letter. Using this value, together with $\gamma=\frac{5}{4}$, in Eq.~(\ref{eq:Zhang3}), one immediately finds
\begin{equation}
\nu^{-1} = \frac{4}{5} (2-\eta) = 1.57.
\end{equation}
Inserting this value for $\nu$ into Eq.~(\ref{eq:Josephson2}) gives the (approximate) nonzero value
\begin{equation}
\alpha = 2-3\nu = 0.09.
\label{eq:approxalpha}
\end{equation}
Earlier, one of us \cite{March:14} has argued that the sum $\alpha+\eta \simeq \frac{1}{8}$, and the results from Eq.~(\ref{eq:approxalpha}) and Wilson's value $\eta\simeq 0.037$ are close to this. Finally, from Eq.~(\ref{eq:HubbardSchofield}), we find that $\delta=4.8$. It seems to the present writers that this change in $\delta$, due to the nonzero value $\alpha=0.09$ estimated in Eq.~(\ref{eq:approxalpha}), from the Zhang prediction $\frac{13}{3}$ in Table~\ref{tab:Zhang} may be the most promising focus for interested experimentalists, though $\eta$ in Eq.~(\ref{eq:Zhang4}) is also sensitive to variations in $\alpha$ near 0.1.

\subsection{Future directions}

While the above discussion has drawn on results from the $\epsilon$-expansion for guidance on the value $\frac{5}{4}$, the Zhang prediction in Table~\ref{tab:Zhang} that has been adopted throughout this Letter for the exponent $\gamma$, we now briefly return to that expansion (see p.~96 in Ref.~\citelow{Kardar:07}). Wilson writes
\begin{equation}
\gamma = 1 + \frac{\epsilon}{6} + \frac{25}{324} \epsilon^2 + O(\epsilon^3).
\label{eq:eps1}
\end{equation}
Also \cite{Kardar:07}
\begin{equation}
\nu = \frac{1}{2} + \frac{\epsilon}{12} + \frac{7}{162} \epsilon^2 + O(\epsilon^3).
\label{eq:eps2}
\end{equation}
Then, it follows almost immediately from Eqs.~(\ref{eq:eps1}) and (\ref{eq:eps2}) that
\begin{equation}
\gamma-2\nu = O(\epsilon^2).
\end{equation}
This result has prompted one of us (NHM) to seek, but now purely by inspection of `known' critical exponents at the integral values $d=2$ (Onsager's solution of the $2d$ Ising model, Ref.~\citelow{Onsager:44}) and $d=4$ (mean field: the leading terms in the $\epsilon$-expansions), a simple formula giving $\gamma-2\nu$ correctly. This is only valid for integral $\epsilon$ therefore, and eventual summation of the $\epsilon$-expansion must intersect these dimensionalities only at integral $d$. The formula advocated by NHM then reads
\begin{equation}
\gamma-2\nu = \frac{2\alpha-1}{d} + \frac{1}{4} .
\label{eq:eps3}
\end{equation}
For $d=3$, it is straightforward to verify that Eq.~(\ref{eq:eps3}) works with $\gamma=\frac{5}{4}$, and $\nu$ given in terms of $\alpha$ by Eq.~(\ref{eq:Josephson2}). In this context, it is well known that $\alpha$ is precisely zero for $d=2$ (Onsager, Ref.~\citelow{Onsager:44}) and $d=4$ (mean field), so that $\gamma-2\nu$ in Eq.~(\ref{eq:eps3}) becomes simply equal to $\frac{1}{4} - \frac{1}{d}$. This latter quantity for $d=2$ is readily verified to lead to Onsager's value for $\gamma-2\nu$, since $\gamma=7/4$ and $\nu=1$ for his case.

Eq.~(\ref{eq:eps3}) is easily verified also for $d=4$, since besides $\alpha=0$, the values $\gamma=1$ and $\nu=\frac{1}{2}$ are indeed the leading terms in the $\epsilon$-expansions. We do not expect that the slope (\emph{i.e.} the derivative with respect to $\epsilon$) at $\epsilon=0$ for $\gamma-2\nu$, which is exactly determined by Eqs.~(\ref{eq:eps1}) and (\ref{eq:eps2}), will agree with Eq.~(\ref{eq:eps3}), which as we have already stressed is true only at integral values 2, 3, and 4 of dimensionality $d$.

Finally, it is to be re-emphasized that the key equations of this section are summarized in Table~\ref{tab:Zhang2}. These lead back precisely to Zhang's predictions in Table~\ref{tab:Zhang}, when the critical exponent $\alpha$ entering the formulae in Table~\ref{tab:Zhang2} is put equal to zero.

\section{Ionization of an electron in an isoelectronic series with $N$ electrons, as the atomic number $Z$ is reduced below that of the neutral atom to a critical value $Z_c (N)$}
\label{sec:iso}

Having genuinely cooperative behaviour in \emph{(i)} melting and magnetic transitions in metals, and \emph{(ii)} near a ferromagnetic critical point as in the insulator CrBr$_3$ or equivalently, because of universality, near a liquid-vapour critical point, we turn to our final, and quite different example. This is about isoelectronic series of atomic ions, such as the He-like case, with $N=2$ and variable atomic number. Again we shall gain insight as to how `self-ionization' comes about by focussing on a solvable model.

For the `real' He series, the electron-electron interaction is of course $u(r_{12}) = e^2 / r_{12}$, where $r_{12}$ denotes the separation between the two electrons. This interaction depends not only on the magnitudes of the electronic position vectors ${\bf r}_1$ and ${\bf r}_2$, but of course also on the angle between them.

\begin{figure}[t]
\centering
\includegraphics[height=0.8\columnwidth,angle=-90]{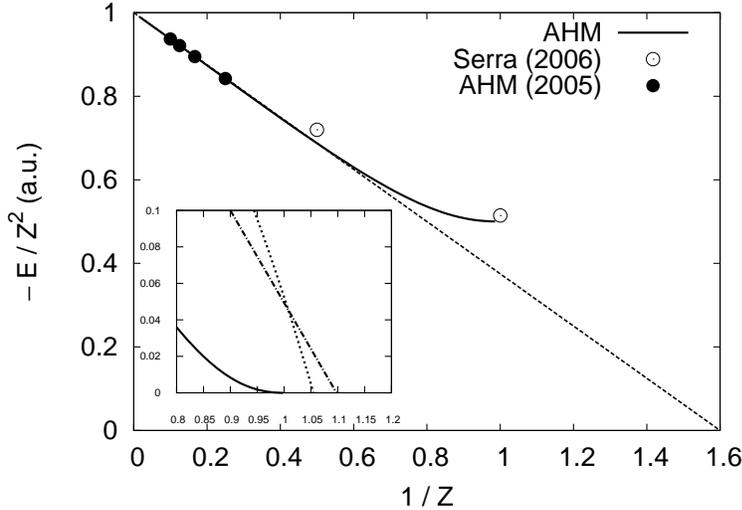}
\caption{Scaled ground-state energy of the AHM model, $-E_{\mathrm{AHM}}/Z^2$, Eq.~(\ref{eq:AHM2}), vs $1/Z$ ($N=2$, solid line, a.u.).
Dashed line depicts tangent line to that curve at $1/Z = 0$.
Open dots at $Z=1$ and $Z=2$ are Serra's \protect\cite{Serra:06} variational values for the ground-state energy of the Temkin model,
whereas solid dots are exact values for the ground-state energy of the AHM model, at $Z=4$, $6$, $8$, and $10$.
Inset shows different critical behaviour of $|E(Z)|/Z^2 - |E(Z_c)|/Z_c^2$ as a function of $1/Z$, as $Z\to Z_c$, in the AHM model (solid line, $Z_c =1$, zero slope), in the Temkin model (dotted line, $Z_c = 0.948072$, weakly infinite slope), and for the $1/r_{12}$ interaction (dashed-dotted line, $Z_c = 0.911028$, finite slope). See Ref.~\protect\citeonline{Glasser:14} for more details.}
\label{fig:AHM}
\end{figure}

This prompted Amovilli, Howard, and March (AHM, below) \cite{Howard:05b,Amovilli:05,Amovilli:08} to generalize the early work of Temkin \cite{Temkin:62} who introduced the so-called $s$-state model of He, by taking the Hamiltonian as
\begin{equation}
H_{\mathrm{T}} = \frac{1}{2} (p_1^2 + p_2^2 ) - Ze^2 \left( \frac{1}{r_1} + \frac{1}{r_2} \right) + \frac{e^2}{r_>} ,
\label{eq:Temkin}
\end{equation}
where $r_> = (r_1 + r_2 + |r_1 - r_2 |)/2$. Then, $H_{\mathrm{AHM}}$ was defined by adding an additional radial correlation $\delta(r_1 - r_2 )$ to Eq.~(\ref{eq:Temkin}), to obtain the model Hamiltonian
\begin{equation}
H_{\mathrm{AHM}} = H_{\mathrm{T}} + \delta(r_1 - r_2 ).
\label{eq:AHM1}
\end{equation}
Then, as derived by AHM, the exact ground-state energy of this model is given by
\begin{equation}
E_{\mathrm{AHM}} (Z,N=2) = -Z^2 +Z - \frac{1}{2} - \frac{6(Z-1)^3}{16Z^2 -25 Z +10} .
\label{eq:AHM2}
\end{equation}
We display in the solid line of Fig.~\ref{fig:AHM} a plot of $-E_{\mathrm{AHM}}/Z^2$ vs $1/Z$ for the ground-state energy of the model
Hamiltonian, Eq.~(\ref{eq:AHM1}). The dashed line shown is the tangent at $1/Z=0$ having slope $-\frac{5}{8}$, as is readily verified
from the exact model ground state energy, Eq.~(\ref{eq:AHM2}). The open dots at $Z=1$ and $2$ are Serra's \cite{Serra:06} variational values 
for the ground-state energy of Temkin's model Hamiltonian, Eq.~(\ref{eq:Temkin}), while the exact energies from Eq.~(\ref{eq:AHM2})
are shown with solid dots at $Z=4$, $6$, $8$, and $10$.

But the `phase transition' focussed on in this section, where a single electron is ionized as $Z$ is reduced well below the neutral
atom value $Z=2$ is readily shown to occur from Eq.~(\ref{eq:AHM2}) at $Z=1$. This, of course, is a model prediction from the exact
Hamiltonian $H_{\mathrm{AHM}}$. Serra \cite{Serra:06} shows variationally that his critical $Z_c$ is 0.948072. For the real He series, with 
interaction $e^2 / r_{12}$, the important early work of Li and Shakeshaft \cite{Li:05} established for this He series that 
$Z_c (N=2) = 0.911028$.

We stress via Fig.~\ref{fig:AHM} the way the three curves referred to behave as the respective $Z_c$ values are approached from above.
For the AHM model, the zero slope at $Z_c=1$ in Fig.~\ref{fig:AHM} follows almost immediately from Eq.~(\ref{eq:AHM2}). Serra's \cite{Serra:06} variational value $Z_c$ quoted above and his factor $(Z-Z_c)^\alpha$ with $\alpha=0.996$ makes the slope nearly infinite at his $Z_c$. For
the He atom series itself, the corresponding curve comes into the critical $Z_c$ with finite slope.

\begin{figure}[t]
\centering
\includegraphics[height=0.8\columnwidth,angle=-90]{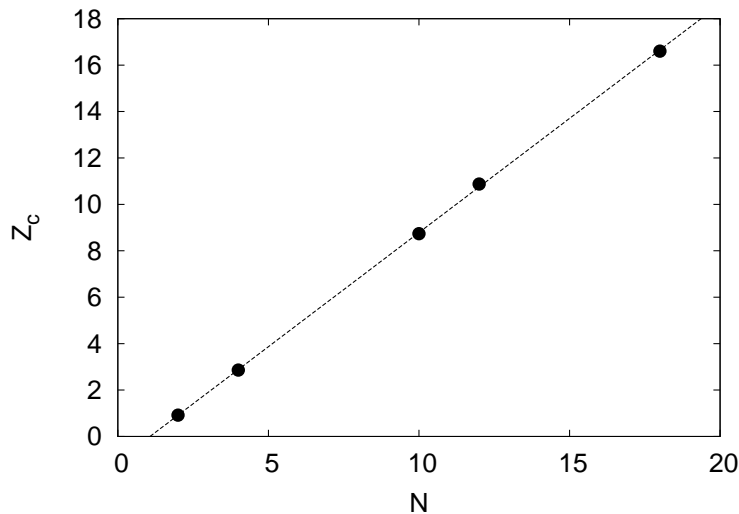}
\caption{Critical atomic numbers $Z_c(N)$, within corrected Hartree-Fock theory \protect\cite{Cordero:13}, at which
a non-relativistic ion is no longer capable of binding $N$ electrons. Dashed line is the best linear fit to the values. (Redrawn after Ref.~\protect\citeonline{Cordero:13}.)}
\label{fig:Cordero}
\end{figure}

We conclude this section by noting that Cordero, March, and Alonso \cite{Cordero:13} report the critical $Z_c (N)$ for real series of atomic ions. First, as these authors note, it has been known for some decades that the value of $Z_c$ obeys the inequalities \cite{Baker:90}
\begin{equation}
N-2 < Z_c < N-1 .
\end{equation}
Cordero \emph{et al.} \cite{Cordero:13} have used available data to plot $Z_c (N)$ vs $N$ over the range $N=2-18$ (Fig.~\ref{fig:Cordero}). Approximate fits are also provided, which turn out to be linear in the number of electrons in the isoelectronic series considered. It is also noteworthy in the present context that Amovilli and March \cite{Amovilli:06c} have carefully studied the long-range asymptotic behaviour of the ground-state electron density $n(r)$ in the He isoelectronic series with $N=2$, as a function of atomic number $Z$. In non-relativistic quantum mechanics, $n(r)$ has the asymptotic behaviour \cite{Hoffmann-Ostenhof:78} $n(r) \varpropto \exp (-2\sqrt{2 I} r)$ for large $r$, where $I$ is the ionization potential. In Ref.~\citeonline{Amovilli:06c}, $I(Z)$ was obtained from diffusion Monte~Carlo calculations, and has the form $I(Z)\simeq 0.218 (Z-Z_c ) + 0.507 (Z-Z_c )^2$. In Ref.~\citeonline{Cordero:13}, Cordero \emph{et al.} generalize this to varying nuclear charge $Ze$.

\section{Concluding comments}
\label{sec:conclusions}

Even given a model Hamiltonian, such for example as that of Hubbard \cite{Hubbard:63} for describing strong electron correlations in narrow energy bands (as in the $3d$ transition metals), it is a highly non-trivial matter to calculate, for instance, the Curie temperature $T_{\mathrm{C}}$. Thus, as both Hubbard \cite{Hubbard:79} and Schrieffer \emph{et al.} \cite{Schrieffer:71} have shown, $T_{\mathrm{C}}$ is at least an order of magnitude too high when the Hubbard $U$ is neglected.

Thus, it seems to us that the heavy rare-earth metals may be exceptional from the viewpoint of melting and magnetic transitions in that Ayuela and March \cite{Ayuela:14} have shown that these are together well characterized by the single quantity $\xi$: the de~Gennes factor. However, having stressed that, it is, as yet only for the N\'eel temperature, $T_{\mathrm{N}}$, that the explicit form in terms of $\xi$, \emph{viz.} $T_{\mathrm{N}} \varpropto \xi^{2/3}$, is well established, and then semiempirically only. However, understanding arising from the de~Gennes factor $\xi$ as the independent variable enabled Ayuela and March \cite{Ayuela:14} (see also their Fig.~2) to construct an `unconventional phase diagram' for the six rare-earth metals reviewed here.

Turning to critical exponents, section~\ref{sec:CrBr3} has set out what should be a final theory of critical exponents for the $3d$ Ising model. This is altogether motivated by the completely analytical predictions of Zhang \cite{Zhang:07a}. However, questions have been raised against the prediction of zero for the critical exponent $\alpha$, though it is surely small ($\simeq 0.1$). Therefore, section~\ref{sec:CrBr3} reviews work of one of us \cite{March:14,March:14a}, in which all critical exponents except $\gamma$ (equal to $\frac{5}{4}$ to within both experimental and theoretical `error') are tabulated in Table~\ref{tab:Zhang2} above. These values are presented solely in terms of $\alpha$. Once this exponent is settled either by computer simulation (most probably) or experiment, all the six exponents will be finally known, given $\gamma=5/4$ as known either exactly, or to high accuracy, as stressed above.

Finally, following these two examples of cooperative phase transitions, a quite different type of transition in isoelectronic series of atomic ions is reviewed, based on the AHM analytical model \cite{Howard:05b,Amovilli:05,Amovilli:08}, but supplemented by diffusion Monte~Carlo simulations on the He series with two electrons but varying atomic numbers.

\section*{Acknowledgements}

NHM wishes to acknowledge that his contribution to this review was made
during a visit to the University of Catania, Italy. He thanks Professors R.
Pucci and G. G. N. Angilella for their kind hospitality, and INFN, Sez. Catania,
for partial financial support. NHM also acknowledges that his continuing affiliation with the University of Antwerp is made possible by the generosity of Professors D.~Lamoen, C.~Van~Alsenoy, and P.~Geerlings.

\section*{References}

\bibliographystyle{ijmpb}
\bibliography{a,b,c,d,e,f,g,h,i,j,k,l,m,n,o,p,q,r,s,t,u,v,w,x,y,z,zzproceedings,Angilella}

\end{document}